# Experimental and theoretical studies of WO$_3$/Vulcan XC-72 electrocatalyst enhanced H$_2$O$_2$ yield ORR performed in acid and alkaline medium

João Paulo C. Moura[1], Lanna E. B. Luchetti[1], Caio M. Fernandes[1], Aline B. Trench[1], James M. Almeida[2], Camila N. Lange[1], Bruno L. Batista[1], Mauro C. Santos[*1]

1 *Laboratório de Eletroquímica e Materiais Nanoestruturados, Centro de Ciências Naturais e Humanas, Universidade Federal do ABC, CEP 09210-170, Rua Santa Adélia 166, Bairro Bangu, Santo André, SP, Brazil*

2 *Ilum, Centro Nacional de Pesquisa em Energia e Materiais (CNPEM). Rua Lauro Vannucci, 1020, Fazenda Santa Cândida, 13087-548, Campinas – SP, Brazil.*

Corresponding Author:
*E-mail:* mauro.santos@ufabc.edu.br

# ABSTRACT

The oxygen reduction reaction (ORR) plays a pivotal role in clean energy generation and sustainable chemical production, particularly in the synthesis of hydrogen peroxide ($H_2O_2$). In this study, $WO_3$/Vulcan-XC72 electrocatalysts have been synthesized and characterized for ORR applications. We assessed the ratio of $WO_3$ to Vulcan-XC72 and investigated the impact of electrolytes pH (covering acidic and alkaline media) on the ORR process. The results revealed that $WO_3$ with a monoclinic crystalline phase and nanoflower-like morphology was successfully synthesized, and confirmed an improvement in surface properties, with an increase in hydrophilicity and superficial oxygenated species. Electrochemical studies showed that $WO_3$/C was the most selective for $H_2O_2$ electrogeneration, compared to pure Vulcan-XC72, in both acidic and alkaline media. These results indicate that the ORR on the $WO_3$/C electrocatalyst surface has a pH-dependent mechanism. Using $WO_3$/C GDEs, an accumulation of 862 mg $L^{-1}$ of $H_2O_2$ was achieved after 120 min of electrolysis at 100 mA $cm^{-2}$. The higher selectivity of $WO_3$/C could be related to the presence of more oxygen functional acid species on the catalyst surface and increased hydrophilicity compared to pure Vulcan, as well as a synergistic effect of the $WO_3$ nanoflowers in ORR, confirmed by theoretical calculations. The results reveal that $WO_3$/Vulcan is a promising catalyst for $H_2O_2$ electrogeneration via the ORR.

KEYWORDS: ORR; DFT; $H_2O_2$ electrogeneration; Carbon; $WO_3$

# 1. Introduction

The oxygen reduction reaction (ORR) is a key process for sustainable energy solutions and environmentally friendly chemical applications. This electrochemical transformation, where oxygen is converted to water or hydrogen peroxide, holds the promise of revolutionizing the landscape of green energy generation and cleaner chemical production.[1–3] One aspect of particular interest is the production of hydrogen peroxide ($H_2O_2$) via two-electron ORR pathway. This versatile and eco-friendly chemical has found applications in pharmaceuticals, wastewater treatment, and chemical synthesis.[2,4]

Carbon materials have gained attention among electrocatalysts for the two-electron ORR pathway, thanks to their remarkable versatility and adjustability. [5] They are widely regarded as one of the most promising alternatives to noble metal-based catalysts. However, it is important to point out that pristine carbon materials inherently display limited catalytic activity for the two-electron ORR. As a result, efforts to enhance their electrocatalytic properties for $H_2O_2$ generation have led to the development of surface modifications or carbon material reconstruction. Several well-documented approaches have been explored, including the creation of porous structures[6,7], the introduction of defects by heteroatom doping[8–11], and surface functionalization.[12,13] The modifications of carbon-based catalysts play a crucial role in enhancing their efficacy. These strategies collectively aim to optimize the electronic characteristics of carbon materials, making them more effective catalysts for selective two-electron electrochemical ORR.

The ORR efficiency, particularly in the context of the $H_2O_2$ generation pathway is influenced by several key factors. Catalyst activity plays a fundamental role, and selecting a given material and tuning its surface properties can significantly impact the reaction yield. Modifying carbon materials with metallic oxides has proven to be a suitable approach for achieving optimum efficiency for the two-electron ORR. In this context, tungsten trioxide ($WO_3$) is a metal oxide with good electrochemical stability, good conductivity, low toxicity, it is relatively abundant in nature, and it is easily synthesized with a relatively low cost.[14–16]

In this paper, we will examine the electrocatalytic behavior of $WO_3$ nanostructures supported on Vulcan XC72. The synthesis and characterization of the $WO_3$/Vulcan XC72 electrocatalyst are explored, aiming to improve its catalytic performance for the ORR. In addition, the pH can have a substantial influence on the ORR efficiency, promoting or hindering the desired reaction pathways, [17–21] which is why understanding the mechanistic pathways of ORR in the context of $H_2O_2$ production is of paramount importance. We investigated the impact of different $WO_3$ to Vulcan XC72 ratios and pH values on the ORR performance, encompassing both acidic and basic media. This research provides novel insight into how the pH can influence the reaction pathway and enhance the catalytic activity of carbon-based electrocatalysts modified with $WO_3$ for the $H_2O_2$ electrogeneration.

## 2. Experimental

### 2.1. Materials

Tungsten (VI) chloride (≥66.5%, Sigma Aldrich) and polytetrafluoroethylene (60wt % dispersion in $H_2O$, Sigma Aldrich), Vulcan XC-72 (Cabot), and ethanol (≥ 99.5, Synth). All the chemical reagents were of analytical grade and used without additional purification treatment.

### 2.2. WO$_3$ synthesis

The WO$_3$ oxide was synthesized by a solvothermal method. Briefly, 0.793 g of WCl$_6$ was dissolved in 40 mL of ethanol and stirred for 30 min. The well-dissolved solution was then transferred to a 50 mL Teflon-lined stainless-steel autoclave and heated at 160 °C for 12 h in a furnace. Following the hydrothermal treatment, the autoclave was cooled down to room temperature, the obtained products were washed with water and ethanol several times, dried in an oven at 70 °C for 6 h, and then the powder was heated at 500 °C for 2 h in a furnace.

### 2.3. WO$_3$/C electrocatalyst preparation

The WO$_3$ electrocatalysts supported on carbon Vulcan XC72 (at 1, 3, and 5% w/w) were prepared by a wet impregnation method[22]. An appropriate quantity of WO$_3$ was added to 0.5 g of Vulcan XC72 and suspended in 30 mL of deionized water under vigorous magnetic stirring for 5 h. The prepared electrocatalysts were then dried in an electric oven at 90 °C.

### 2.4. Electrocatalyst characterization

X-ray powder diffraction (XRD) measurements were performed in transmission geometry, using the conventional diffractometer model STADI-P (Stoe®, Darmstadt, Germany) operating at 40 kV and 40 mA and equipped with a primary beam monochromator of Ge(111), providing CuKα1 radiation ($\lambda$ = 1.54056 Å). The powdered samples were deposited between two sheets of cellulose acetate, and the sample holder was kept rotating during the data collection. The integrated intensities were recorded by a linear detector model Mython 1K (Dectris®, Baden, Switzerland) in 10°– 90° 2θ range at a scan rate of 2° min$^{-1}$. Raman scattering was applied using a Horiba-Jobin-Yvon, model T64000 coupled to a laser light at 532 nm and applying an exposure time of 40 s. Material morphologies and microstructures were investigated using a FESEM JEOL JSM-7401 operating at 30 kV and an HRTEM JEOL JEM 2100 microscope operating at 200 kV.

Wettability was characterized through electrocatalyst contact angle measurements using a goniometer (GBX Digidrop). Briefly, 1 mg/mL dispersions (catalyst powder/deionized water) were prepared through sonication for 1 min using a tip ultrasonicator. A 40 µL aliquot of each dispersion was then deposited on a vitreous carbon plate and dried to form a thin and homogeneous film. Subsequently, 5 µL of Milli-Q water, or electrochemical electrolytes were dropped on the film surface to determine the contact angle. Measurements were obtained in triplicate using the Windrop software.

**2.5. Oxygen reduction reaction study**

Electrochemical measurements were performed employing an Autolab PGSTAT 302 N Potentiostat/galvanostat with a rotating ring-disc electrode system (RRDE) (Pine Instruments) using a 125 mL electrochemical cell. A 2 cm$^2$ platinum counter electrode (Pt), reference Hg/HgO or Ag/AgCl electrodes, and a working electrode composed of a GC disc (area = 0.2475 cm$^2$) and gold (experimental collection factor of N = 0.28) or platinum ring (N = 0.21) both with area = 0.1866 cm$^2$ were employed. 1 mol L$^{-1}$ NaOH (pH 14) or 0.1 mol L$^{-1}$ K$_2$SO$_4$ (adjusted pH 3 with H$_2$SO$_4$) were used as the supporting electrolytes for the electrochemical measurements. The applied potentials in the reference electrodes were adjusted to that of a reversible hydrogen electrode (RHE). Electrochemical impedance spectroscopy (EIS) was measured within the frequency range from 10$^{-5}$ to 10$^{-1}$ Hz and an amplitude of 10 mV with 10 points per decade, and the polarization potential was set at the diffusion limit region.

The electrocatalysts were deposited on the glassy carbon disc working electrode as thin, porous layers by drop casting. Dispersions containing 1 mg/mL of the electrocatalyst in water were homogenized using a tip ultrasonicator. After homogenization, 20 µL of the dispersions were deposited on the disc electrode surface. After drying, 20 µL of a 1:100 Nafion solution (v/v, Nafion: deionized water) was pipetted on the film covering the working electrode and dried. The electrolyte was previously saturated with oxygen for 30 min for all electrochemical analyses, maintaining the same flow throughout all electrochemical measurements, performed in duplicate at a scan rate of 5 mV s$^{-1}$ at room temperature.

The number of transferred electrons in the ORR and the catalyst selectivity toward $H_2O_2$ production were determined based on Equations 1-2:

$$X_{H_2O_2} = \frac{2I_r/N}{-I_d + I_r/N} \quad (1)$$

$$n_t = 2[(X_{H_2O}) + 1] \quad (2)$$

where $I_r$, $I_d$ and $N$ are the ring current, disk current and collection efficiency of the RRDE, respectively.

## 2.6. In-Situ H$_2$O$_2$ electrosynthesis

In situ $H_2O_2$ electrosynthesis was accomplished using a gas diffusion electrode (GDE) cathode prepared by a hot-pressing procedure using 1% $WO_3$/Vulcan XC-72 containing 20 % (w/w) of PTFE dispersion, as described previously by our research group [12,13,23]. Electrogeneration essays was carried out in an undivided electrochemical cell using a reference Ag/AgCl electrode, a Pt counter electrode (5 cm$^2$) and the fabricated GDE (3.5 cm$^2$) with continuous $O_2$ supplying at 0.2 bar. The electrolyte comprised 350 mL of 0.1 mol L$^{-1}$ $K_2SO_4$ (pH 3 adjusted with $H_2SO_4$). The electrolysis process was performed by chronopotentiometry technique applying 50, 75, and 100 mA cm$^{-2}$ for 120 min. The $H_2O_2$ accumulation was quantified colorimetrically using $(NH_4)_6Mo_7O_{24} \cdot 4H_2O$ ($2.4 \times 10^{-3}$ mol L$^{-1}$) in $H_2SO_4$ (0.5 mol L$^{-1}$) in a spectrophotometer Varian Cary 50.

**2.7. DFT calculations**

DFT calculations were performed with the PWscf code implemented in the Quantum ESPRESSO suite.[24] Core electrons were described pseudopotentials[25] and valence electrons with plane wave functions and a 60 Ry kinetic energy cutoff. GGA-PBE functionals were employed to address the exchange-correlation energy. [26] The unit cell lattice parameter was optimized and then replicated to build slab models, periodic in two directions and with a 20 Å vacuum layer in the perpendicular axis, to correctly describe the catalytic surface. The slabs were optimized until a convergence threshold on the forces per atom of $10^{-3}$ a.u. and $10^{-4}$ a.u. on total energy were both reached, with half of the bottom layers coordinates kept fixed while the upper half of superficial atoms were able to relax. Following what was observed with XRD analysis, the low Miller surfaces have been selected to determine the role of surface orientation on the catalytic activity, namely (001), (010), and (100). The Hubbard parameter of $U = 3.4788$ $V$ was determined *ab initio*, according to what has been proposed by Matteo Cococcioni and Stefano de Gironcoli [27] and implemented in the PWscf code suite.[24] The 2-electron ORR was investigated with the computational hydrogen electrode (CHE) framework. [28]

**3. Results and discussion**

**3.1. WO$_3$ nanostructure characterizations**

The crystalline structure of WO$_3$ was investigated with XRD analysis, shown in **Figure 1**. The WO$_3$ XRD pattern spectrum indicates good crystallinity with phase purity. All the diffraction peaks are coincident with a monoclinic structure as according to the Inorganic Crystal Structure Database (ICSD) card no. 17003 (monoclinic structure with a space group P21/n). No peaks belonging to any other phase were identified, demonstrating the high purity of the products. The WO₃/Vulcan XC72 patterns at different loadings are also shown in **Figure 1**. The broad peak at 2θ ~ 24º is assigned to a typical amorphous carbon structure related to Vulcan XC72. Furthermore, the $WO_3$ signals were observed, and the intensity peaks increased with the metallic oxide loading, indicating the successful impregnation of $WO_3$ onto Vulcan XC72.

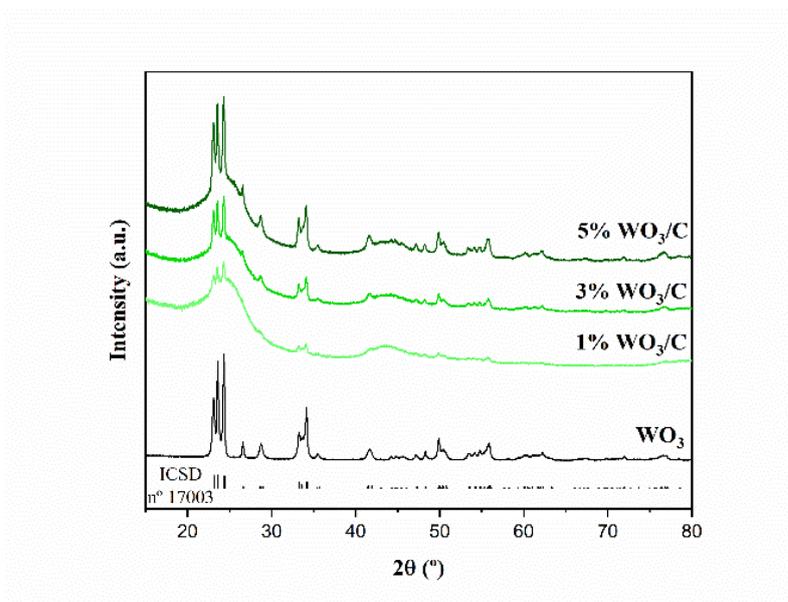

**Figure 1.** XRD patterns of the solvothermally synthesized $WO_3$ and prepared $WO_3$/C electrocatalysts. The vertical sticks below are relative peak positions and intensities using ICSD cards as reference.

The $WO_3$ morphology and nanostructure were investigated by FESEM and HRTEM shown in **Figure 2**. It can be seen in **Figure 2a** that large-scale flower-like nanostructures with an average size ranging between 0.6–1.6 μm in diameter were obtained. **Figure 2b** shows several dozen nanosheets (petals) connected to each other to form 3D nanoflowers by self-assembly. It can be noted that the synthesized $WO_3$ sample showed high crystallinity. Fast Fourier transform (FFT) was performed on the selected area in yellow (**Figure 2c**) and indicated the presence of monoclinic $WO_3$ based on the

(020) plane with an interplanar distance of 3.70 Å (ICDS-17003) [29,30] and the (022) plane with an interplanar distance of 2.70 Å.[31] These results corroborate the results observed in XRD and show that the WO$_3$ sample was successfully synthesized.

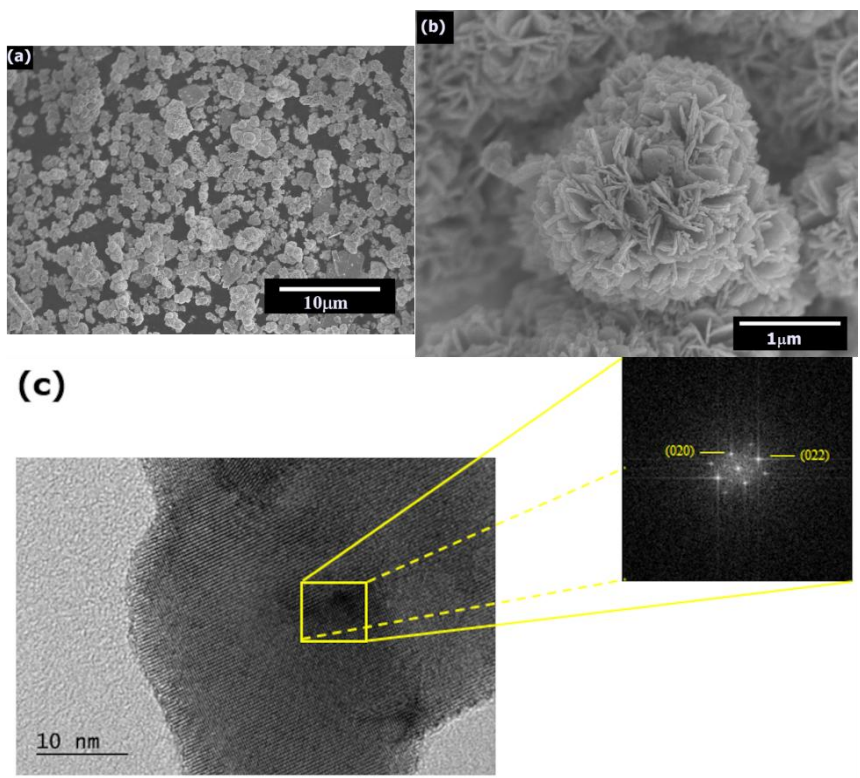

**Figure 2. (a,b)** FESEM and **(c)** HRTEM images of the synthesized WO$_3$.

**Figure 3a.** shows the high-resolution C 1s XPS spectra for 3 % WO$_3$/C (the other catalysts are shown in **Figure S1**). The spectra were deconvoluted into five component peaks which are related to surface functional groups: graphitic sp$^2$ carbon (284.8 eV), sp$^3$ hybridized carbon (286.5 eV), C-O bonds in hydroxyls and/or epoxides species (288.1 eV), C=O bonds in carbonyls and/carboxyl species groups (289.6 eV), and π -π* plasmon transitions (290.9 eV).[32,33] This analysis revealed an increase in superficial oxygenated species groups at Vulcan XC72 modified with WO$_3$ in comparison with pure carbon. These oxygenated species can improve the mass transfer process between the catalyst surface and the dissolved oxygen. Previous theoretical works in the literature indicate that oxygen functional groups enhanced the two-electron ORR selectivity of carbon surfaces.[34,35] Traditionally, the assessment of ORR electrocatalysts has often been limited to measuring contact angles solely with Milli-Q water, providing insights into the surface hydrophilicity. However, this approach overlooks a critical aspect of real-world applications, where the electrocatalysts interact with specific electrolytes. In this context, our study underscores the importance of measuring contact angles not only with water but also with the actual acidic and basic electrolytes used in the system.

The water dropped test in milli-Q water shown in **Figure 3b** reveals an increase in hydrophilicity properties with the carbon-supported $WO_3$ when compared to pure Vulcan XC72, indicated by a lower contact angle. The observed values for $WO_3/C$ decrease in the alkaline and acidic electrolytes, as also observed in **Figure 3b**. The contact angle of a liquid droplet on a solid surface is influenced by the intermolecular interactions between the liquid, solid, and gas phases. So, the contact angle of a 1 mol $L^{-1}$ NaOH (pH 13) solution on a $WO_3/C$ electrocatalyst is higher than the contact angle of a 0.1 mol $L^{-1}$ $K_2SO_4$ (adjusted to pH 3 with $H_2SO_4$) solution, due to differences in surface chemistry and wettability. [36]

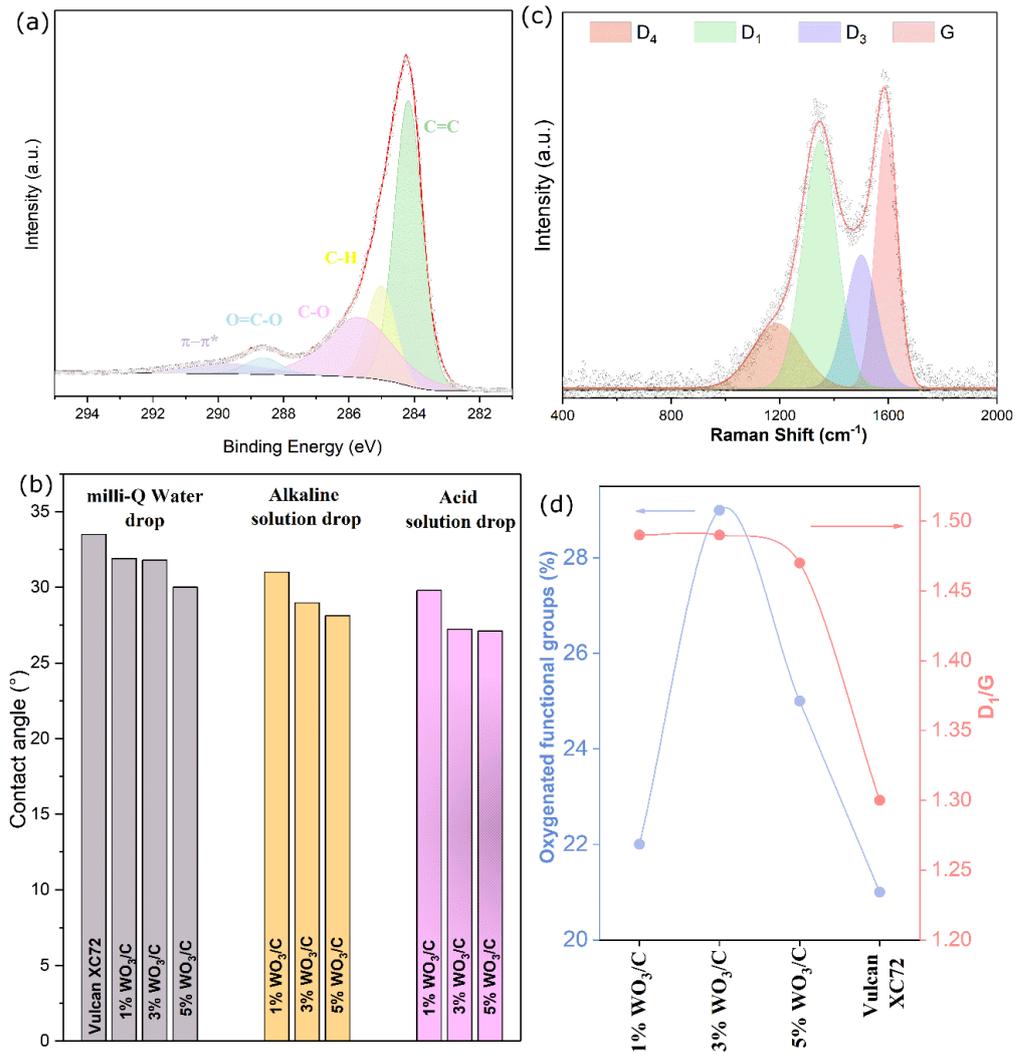

**Figure 3.** **(a)** High-resolution C 1s XPS profile 3% WO$_3$/C, and **(b)** electrocatalyst contact angles measurement, **(c)** 3% WO$_3$/C Raman spectra, and **(d)** relationship between oxygenated functional content and D1/G Raman defects.

At the NaOH solution (pH = 13), the medium contains negatively charged hydroxide ions (OH$^-$). The WO$_3$/C surface might also have negative charge sites. The electrostatic repulsion between the negatively charged surface and the hydroxide ions tends to make the droplet more rounded, thus increasing the contact angle. At a lower pH of 3, the K$_2$SO$_4$ solution would be acidic (even after adjusting the pH). The presence of hydrogen ions (H$^+$) at the surface might neutralize some negative charges. Additionally, sulfate ions (SO$_4^{2-}$) from the salt could interact differently with the WO$_3$/C surface, possibly reducing repulsion. Consequently, the contact angle in this case is lower due to reduced electrostatic repulsion between the droplet and the surface.[37]

This hydrophilicity profoundly impacts oxygen gas ($O_2$) transport and the adsorption process at active sites, thereby improving the ORR activity. Higher hydrophilicity not only enhances $O_2$ transfer dynamics but also facilitates more efficient proton migration to the gas-liquid-solid interface. This multifaceted enhancement highlights the pivotal role of surface wettability in the overall electrocatalytic performance. [38–40]

Finally, Raman spectroscopy was performed to corroborate the aforementioned results. **Figure 3c** shows the 3% $WO_3$/C spectrum (the other spectra are shown in **Figure S2**). Two distinct peaks observed at approximately 1340 and 1600 cm$^{-1}$ were deconvoluted into four subcomponents using Gaussian fitting to accurately assess the disorder and carbon defects. According to the literature, the $D_1$ band is associated with defects attributed to the vibrations of sp$^3$ carbon atoms, while the G band is generated as a result of highly ordered sp$^2$-bonded carbon atoms.[41,42] The $D_3$ band arises from out-of-plane vibrations caused by defects and heteroatoms, which ultimately vanish during graphitization. The $D_4$ band is associated with CH species in aliphatic hydrocarbon chains.[43]

The $D_1$/G area ratio represents an overall proportion between the disordered and graphitic carbon structure of the materials. The $D_1$/G ratios of $WO_3$/C electrocatalysts lie around 1.49, indicating that there are more defects in the surface carbon layer compared with raw Vulcan XC72, caused by the $WO_3$–Vulcan XC72 interaction. Previous works reveal that defective carbon structures were closely related to an improvement in two-electron ORR selectivity. [23,44] **Figure 3d** shows the relationship between carbon structural defects $D_1$/G and oxygenated functional groups at the catalyst surface determined from XPS. The results demonstrate that the interaction between $WO_3$ and Vulcan XC72, particularly in small oxide:carbon ratios, enhances its surface properties, and consequently its effectiveness as an ORR electrocatalyst.

### 3.2. Electrochemical measurements

The ORR performance of the synthesized catalysts was further assessed using the rotating ring-disk electrode (RRDE) polarizations in $O_2$-saturated electrolytes. As shown in **Figure 4a-b**, the collected linear scanning voltammetry (LSV) plots on the ring and disk electrodes for the $WO_3$/C in alkaline and acid electrolytes respectively. The ORR activity and $H_2O_2$ selectivity were evaluated based on disk and ring currents, respectively. The calculated $H_2O_2$ selectivity and electron-transfer number (n) ascertained from ring and disk current are also shown in **Figure 4c-d** as functions of the applied potential.

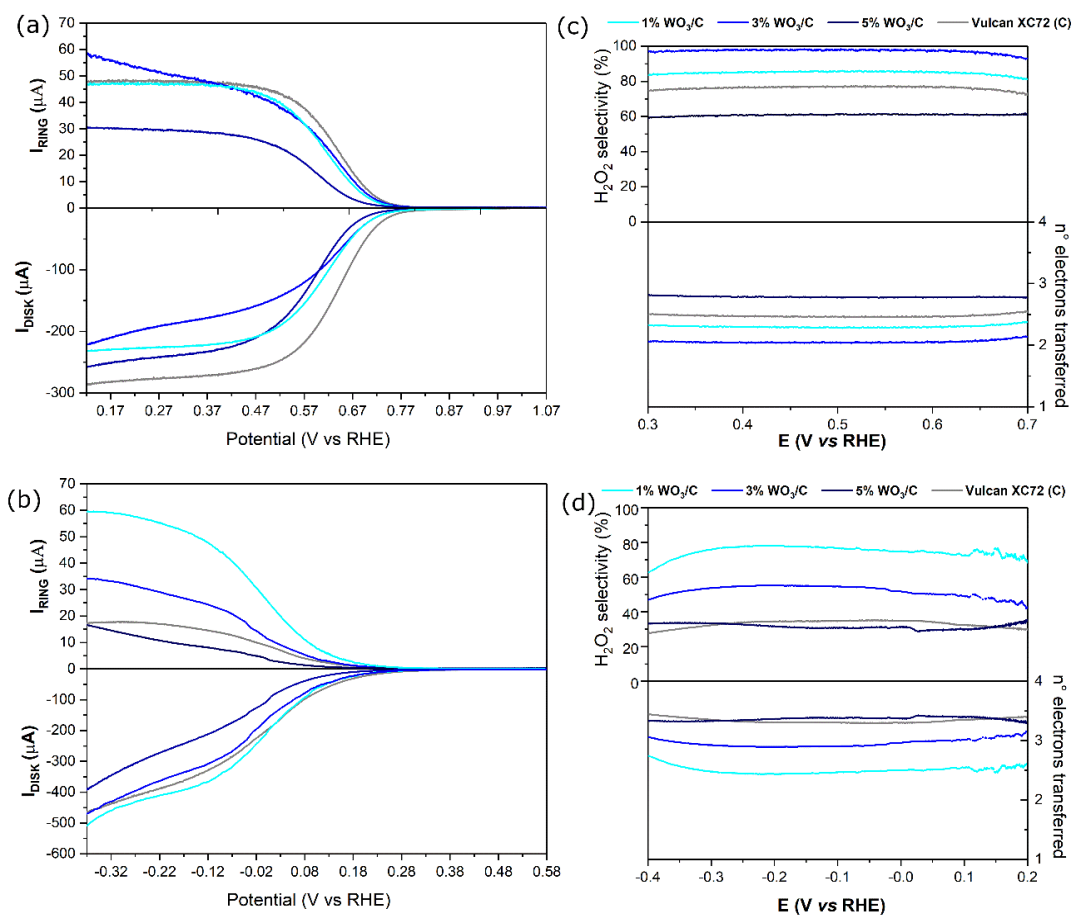

**Figure 4.** Steady-state polarization curves for the ORR employing WO$_3$/Vulcan XC72 and pure Vulcan XC72, indicating the electron transfer number and H$_2$O$_2$ selectivity (%) in **(a,c)** alkaline electrolyte and **(b,d)** acid electrolyte.

It can be seen by the LSV performed in the alkaline medium that the modified catalyst showed a slight decrease in ORR activity in this case. This is indicated by the disk current relative to Vulcan XC72, but this trend is not reflected in the selectivity for the two-electron ORR pathway. The 3% WO$_3$/C showed the highest selectivity of near 100% at a large potential range and with ~ 2.1 electrons transferred. In the acid medium, the ORR performance of the WO$_3$/Vulcan catalysts and the two-electron ORR selectivity is better than pure Vulcan XC72. For instance, the 1% WO$_3$/C catalyst reached 80% selectivity in this case. It also can be observed that the catalyst had more positive H$_2$O$_2$ onset potential in the alkaline medium than it did in the acid medium (detailed in **Figure S3**). Hence, the modification of WO$_3$/C was enhanced for the two-electron ORR in the alkaline electrolyte, compared to the acidic medium.

These results suggest that the ORR in the acidic electrolyte proceeds in a different catalytic mechanism compared to that in the alkaline medium. Some other works in the literature shed light on pH dependence, for instance, Rojas Carbonel *et al.* reported that the ORR mechanism is influenced by the attraction of protons and hydroxyl ions to various functional groups found on the catalytic surface.[17] A study conducted by Zhao's group revealed the potential and pH-dependent selectivity are linked to the proton's affinity for either the initial or subsequent oxygen in *–O–OH on the carbon catalyst. The proton's preference under acidic conditions is for the subsequent oxygen, leading to decreased H$_2$O$_2$ selectivity.[3]

In order to better understand the kinetics and ORR mechanisms, Tafel plots and electrochemical impedance spectroscopy (EIS) were carried out. The Tafel slopes for the alkaline medium are shown in **Figure 5a**. The values laid around ~ -60 mV dec$^{-1}$ [23], indicating a fast electron transfer followed by a rate-determining step in the chemical process. The ORR rate is primarily determined by how quickly protons species (H$^+$) are transferred between the electrode surface and the adsorbed oxygen.[45] In alkaline medium the H$_2$O molecule serve as proton species source, thus, the more hydrophilic sites representing higher ORR 2e- selectivity. In this case, the proton species transfer step in a non-coupled proton-electron transference is the rate-determining step. This behavior may be associated with the selective adsorption of hydroxyl ions onto the catalyst surface at high pH values. This introduces an additional electron transfer barrier,[17] as evidenced by the EIS spectrum in **Figure 5b**, which reveals resistive electron transfer behavior in an alkaline environment. This suppressing fast ORR kinetics proved favorable for the ORR 2e- pathway.[46]

In an acidic environment, the Tafel slope values were around -180 mV per decade. Among these values, the 1% WO/C catalyst exhibited a lower slope, indicating better kinetic performance compared to the others. Sara Kelly *et al.* investigated a microkinetic model for the ORR on Au (111) surface from both theoretical and experimental approaches. The authors obtained Tafel slope values consistent with our results and correlated those to an unfavorable *O$_2$ adsorption on the surface that contributed to the overall barrier for the first *O$_2$ protonation.[47] These results support that ORR has a pH-dependence mechanism on the WO$_3$/C electrocatalyst surface and in the acid medium, a proton-coupled electron transfer (PCET) was the rate-determining step.[48–51] In acidic electrolytes, the availability of H$^+$ species played an important role in the ORR kinetics, however, the high proton availability and transport to O$_2$ reduction can lead to a 2+2e- mechanism, hindering charge delocalization on the catalyst surface and decreasing the H$_2$O$_2$ yield.[52,53] Additionally, the EIS (**Figure 5d**) reveals lower semicircle values in acidic media, which means a lower charge transfer resistance and a facilitated electron transfer reaction. Therefore, the EIS is consistent with a greater number of electrons being transferred in the ORR

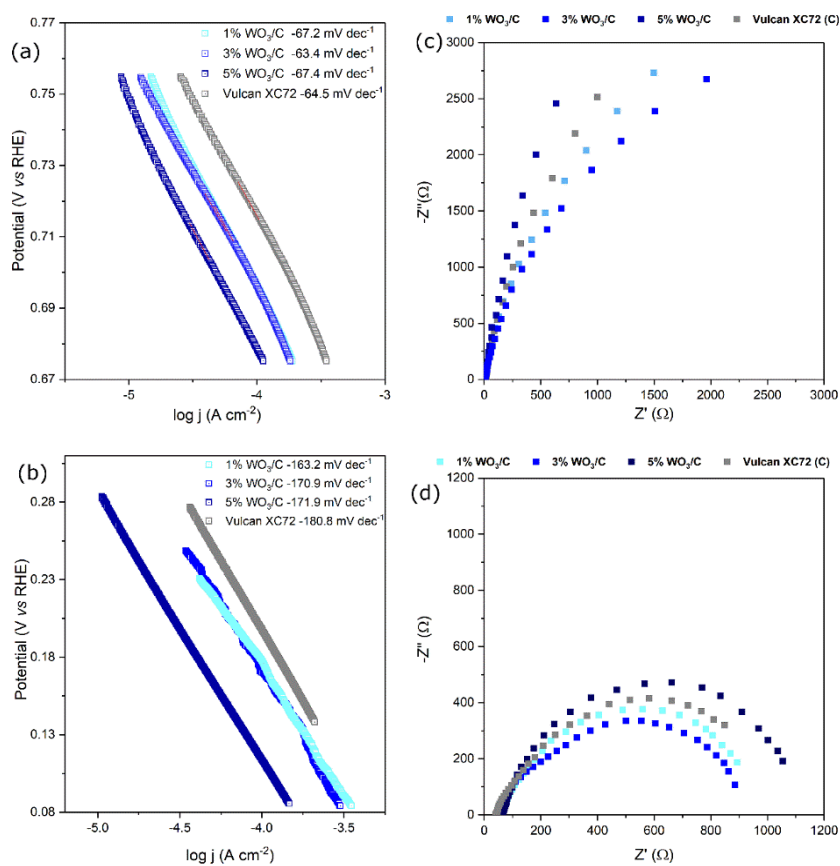

**Figure 5**. Tafel and EIS analysis of WO$_3$/C electrocatalysts in **(a,c)** alkaline medium and **(b,d)** acid medium.

Furthermore, in relative terms, 3% WO$_3$/C showed a better ORR electron transport profile from Nyquist plots (lower charge transfer resistance) for both acid and alkaline media, which can be related to more oxygen functional groups on the surface, identified by XPS analysis, the high content of active oxygenated species groups can act promoting the ORR. [54,55] Finally based on our results, we confirmed that the mechanism changed depending of the medium pH for the same electrocatalyst. In the acid and alkaline medium, 1% WO$_3$/C and 3% WO$_3$/C yielded the best results, respectively, suggesting that the ORR mechanism was influenced by pH, surface chemical state, and possibly by different active site profiles, governed by the oxygenated species content.

The in-situ electrogeneration of $H_2O_2$ is recognized as a highly intriguing technology within the realm of electrochemical advanced oxidation processes (EAOPs), particularly in processes based on electro-Fenton. Several works report that high electrogeneration of $H_2O_2$ is essential for the efficient degradation of persistent pollutants [56,57];It is also known that electro-Fenton applications are more effective in acidic pH. [58,59] In spite of that, we investigated the $H_2O_2$ accumulation through electrolysis using gas diffusion electrode (GDE) based on 1% $WO_3$/C in acid medium (shown in **Figure 6a**). The $H_2O_2$ accumulation for the $WO_3$/C GDE after 120 minutes was 501 mg $L^{-1}$, 682 mg $L^{-1}$, and 862 mg $L^{-1}$ at 50, 75, and 100 mA $cm^{-2}$ density current supplied, respectively. The Current Efficiency (CE) and the energy consumption for $H_2O_2$ electrogeneration were calculated (shown in **Figure 6b**). The applied current density had direct influence in CE, energy consumption, and $H_2O_2$ production. The accumulation of $H_2O_2$ increased due to the greater electrons supply, but at high current supply, parallel reactions occur, including $H_2O_2$ reduction, $H_2$ evolution, and the four-electron ORR pathway. These reactions can take place, decreasing the CE and increasing energy consumption.[23,57] Nevertheless, the $WO_3$/C GDE exhibited excellent values for CE, 80% around, especially when compared with some catalysts reported in literature (shown in **Figure 6c**), this further highlights this catalyst in in-situ $H_2O_2$ production applications using GDE, as well as applications in EAOPs.

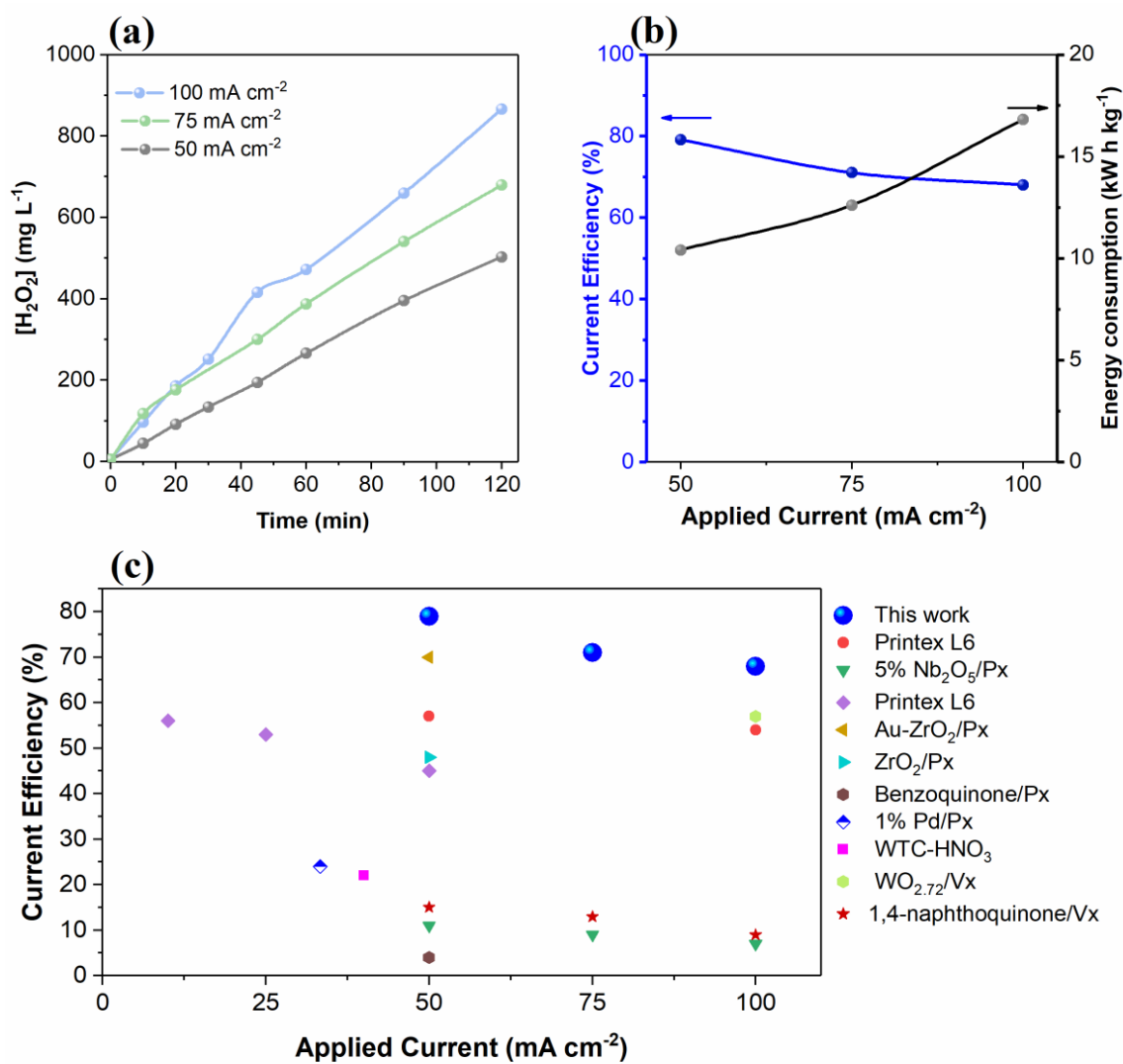

**Figure 6.** **(a)** $H_2O_2$ in-situ accumulation, **(b)** current efficiency and energy consumption for 1% $WO_3$/Vulcan XC72 at different applied currents. **(c)** Current efficiency comparison with some reported works in the literature (references detailed in **Table S1**).

### 3.3. DFT calculations

To obtain further insights regarding the role of $WO_3$ surface orientation and pH effects, DFT calculations were carried out focused on the two-electron ORR pathway, as the electrochemical measurements indicated that the catalysts selected for this study had high activity and selectivity for this mechanism. The results indicate that the intermediate species *OOH effectively binds to the catalytic surface in all cases, while at the same time retaining the O-O bond which is detrimental to promoting the reaction through the two-electron mechanism. [60] The O-O bond length increases during this process from the initial length of 1.21 Å to 1.34 Å, 1.35 Å, and 1.48 Å on the (001), (010), and (100) surfaces, respectively, towards the peroxide O-O distance of 1.48 Å.[61] Therefore, the selected $WO_3$ surfaces yield the expected and desirable catalytic behavior for the two-electron ORR. However, besides an energetically favorable adsorption, the ideal catalyst should also be able to promote the reaction with a low theoretical overpotential.

To order to investigate the process' thermodynamics and determine the theoretical overpotential on $WO_3$ surfaces, the reaction Gibbs free energy diagrams were obtained with the CHE model. The two-electron ORR is a two-step process and within this model, the *OOH intermediate Gibbs free energy - $\Delta G$(*OOH) - can be employed as the activity descriptor. An adequate adsorption/desorption ability of the *OOH intermediate is a key point regarding the selectivity of the oxygen reduction reaction. Ideally, this value should lie in the range of 4.2±0.2 eV for an optimal catalytic activity and selectivity for this mechanism.[62] **Figure 7** shows the Gibbs free energy diagrams for this reaction on $WO_3$ (001), (010), and (100) surfaces. At an open-cell circuit, shown in **Figure 7a**, and indicated by $U = 0$ V, $\Delta G$(*OOH) values are 4.65 eV, 4.45 eV, and 2.89 eV for the different crystalline planes (001), (010), and (100), respectively. While (001) and (010) surfaces lie very close to the ideal $\Delta G$, (100) adsorbs the *OOH intermediate too strongly, yielding a very negative $\Delta G$(*OOH) and consequently, the $H_2O_2$ formation is upwards. This indicates that (100) $WO_3$ planes will be prone to surface poisoning and have poor catalytic activity for this reaction. [63] This observation is in agreement with previous works in the literature. For example, Hurtado-Aular *et al* [64] observed that water molecules adsorbed strongly and with more negative energy values on (100) surfaces than on (001) surfaces of monoclinic $WO_3$. This higher reactivity and stronger bonds with oxygenated species can be explained by the superficial tungsten atoms' coordination: while on (100) each tungsten atom is coordinated to 4 oxygen atoms, on (001) and (010) the coordination number is 5. Furthermore, the (001) and (010) crystalline planes have been previously selected for their performance as gas sensors, photocatalysts, and other applications, therefore corroborating these results. [65,66]

At the thermodynamical limit when $U = 0.7$ V, the theoretical overpotential values can be determined. Those are indicated by the colored bars in **Figure 7b**. It can be seen that the reaction will be unfeasible on the (100) surface, with a very high overpotential of $\eta = 1.33$ V. However, the theoretical overpotential becomes as low as $\eta = 0.23$ V on the (010) surface and $\eta = 0.43$ V on the (001) surface. If these values are corrected to account for the experimental alkaline medium (pH = 13) by subtracting the following term on the Gibbs free energy variation: $G(pH) = kT \times \ln 10 \times pH$ [67,68] the theoretical overpotentials become $\eta = 0.10$ V and $\eta = 0.30$ V, respectively. This implies that both (001) and (010) surfaces will have a shift in the catalytic activity towards the top of the volcano plot, thus improving their performance, as can be seen in **Figure 7c**. These results shed light on the pH influence on the ORR onset potential, indicating that lower values are observed on $WO_3$ sites in an alkaline medium, leading to a higher selectivity for the two-electron pathway, aligned with the experimental results (**Figure S3**).

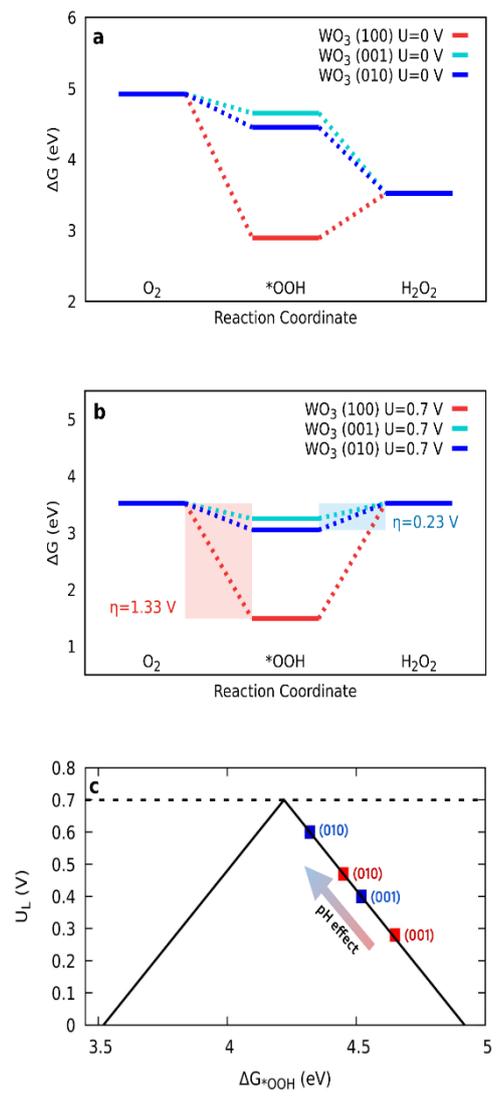

**Figure 7.** Free energy diagram for the oxygen reduction to $H_2O_2$ at $U= 0\ V$ **(a)** and at the equilibrium potential $U= 0.7\ V$ **(b)**. The colored bars indicate the theoretical overpotential obtained for different $WO_3$ crystalline planes. Volcano plot for the 2-electron ORR, where the red indicates the reaction limiting potentials at pH = 0 and blue indicates at pH = 13 **(c)**.

## 4. Conclusions

In conclusion, the synthesis of $WO_3$ flower-like nanostructures using the solvothermal method was successfully achieved, and these nanostructures were employed to modify commercial Vulcan XC72. Analysis through XPS, contact angle measurements, and Raman spectroscopy confirmed an improvement in surface properties, indicating that the modifications tuned the ORR activity and selectivity for the two-electron pathway. This was further confirmed by RRDE polarization analysis. Additionally, this study explores the impact of electrolyte pH on ORR selectivity, with results indicating that the electrolyte pH influences the reaction mechanism. Moreover, theoretical calculations reveal that (001) and (010) $WO_3$ surfaces are catalytically active toward the two-electron ORR and that, in both cases, the basic pH contributes to lower theoretical overpotentials and improved activity. The gas diffusion electrode with 1% WO3/C exhibited excellent performance in the in-situ electrogeneration of $H_2O_2$ in acidic media, accumulating 862 mg $L^{-1}$ after 120 minutes of electrolysis. Ultimately, experimental and computational results demonstrate that the $WO_3$/Vulcan XC72 catalyst exhibits increased selectivity for hydrogen peroxide electrogeneration from ORR, both in acidic and alkaline media. This study offers valuable insights into enhancing the oxide modifying carbon as catalyst for efficient 2-electron ORR in different pH media. It can serve as a helpful guide for optimizing applications related to in-situ hydrogen peroxide electrogeneration.

**CRediT authorship contribution statement**

João Paulo C. Moura Conceptualization, Investigation, Formal analysis, Data curation, and Writing - original draft; Lanna E. B. Luchetti and James M. Almeida Formal analysis and Writing – review & editing; Caio M. Fernandes, Aline B. Trench, Camila N. Lange and Bruno L. Batista worked in Validation and Writing – review & editing; Mauro C. Santos Writing - review & editing, supervision and funding acquisition.

**Declaration of competing interest**

The authors declare that they have no known competing financial interests or personal relationships that could have appeared to influence the work reported in this paper

**Acknowledgments**

The authors wish to thanks to the financial support of the following Brazilian Research Funding Institutions: Fundação de Amparo à Pesquisa do Estado de São Paulo (FAPESP, 2021/05364-7, and 2017/10118-0, 2021/14394-7, 2022/10484-4), Coordenação de Aperfeiçoamento de Pessoal de Nível Superior (CAPES) and CNPq (303943/2021-1, and 150840/2023-3)

**References**

[1] Ren X, Dong X, Liu L, Hao J, Zhu H, Liu A, et al. Research progress of electrocatalysts for the preparation of $H_2O_2$ by electrocatalytic oxygen reduction reaction . SusMat 2023;3:442–70. https://doi.org/10.1002/sus2.149.

[2] Peng W, Tan H, Liu X, Hou F, Liang J. Perspectives on Carbon-Based Catalysts for the Two-Electron Oxygen Reduction Reaction for Electrochemical Synthesis of Hydrogen Peroxide: A Minireview. Energy & Fuels 2023. https://doi.org/10.1021/acs.energyfuels.3c02732.

[3] Zhao X, Liu Y. Origin of Selective Production of Hydrogen Peroxide by Electrochemical Oxygen Reduction. J Am Chem Soc 2021;143:9423–8. https://doi.org/10.1021/jacs.1c02186.

[4] Zhao Q, An J, Wang X, Li N. In-situ hydrogen peroxide synthesis with environmental applications in bioelectrochemical systems: A state-of-the-art review. Int J Hydrogen Energy 2021. https://doi.org/10.1016/j.ijhydene.2020.05.227.

[5]  Petsi P, Plakas K, Frontistis Z, Sirés I. A critical assessment of the effect of carbon-based cathode properties on the in situ electrogeneration of H2O2. Electrochim Acta 2023;470:143337. https://doi.org/10.1016/J.ELECTACTA.2023.143337.

[6]  Rechotnek F, Fragal EH, Galdioli Pellá MC, Fragal VH, Silva R. Enhancement of selectivity towards the synthesis of hydrogen peroxide by dimensional effect in mesoporous carbon. Microporous and Mesoporous Materials 2022;333. https://doi.org/10.1016/j.micromeso.2022.111741.

[7]  Sun L, Sun L, Huo L, Zhao H. Promotion of the Efficient Electrocatalytic Production of H2O2 by N,O- Co-Doped Porous Carbon. Nanomaterials 2023;13. https://doi.org/10.3390/nano13071188.

[8]  Zeng S, Wang S, Zhuang H, Lu B, Li C, Wang Y, et al. Fluorine-doped carbon: A metal-free electrocatalyst for oxygen reduction to peroxide. Electrochim Acta 2022;420. https://doi.org/10.1016/j.electacta.2022.140460.

[9]  Wan J, Zhang G, Jin H, Wu J, Zhang N, Yao B, et al. Microwave-assisted synthesis of well-defined nitrogen doping configuration with high centrality in carbon to identify the active sites for electrochemical hydrogen peroxide production. Carbon N Y 2022;191:340–9. https://doi.org/10.1016/j.carbon.2022.01.061.

[10] Xu J, Cui Y, Wang M, Chai G, Guan L. Pyrimidine-assisted synthesis of S, N-codoped few-layered graphene for highly efficient hydrogen peroxide production in acid. Chem Catalysis 2022;2:1450–66. https://doi.org/10.1016/j.checat.2022.04.011.

[11] Wang X, Zhang Q, Jing J, Song G, Zhou M. Biomass derived S, N self-doped catalytic Janus cathode for flow-through metal-free electrochemical advanced oxidation process: Better removal efficiency and lower energy consumption under neutral conditions. Chemical Engineering Journal 2023;466. https://doi.org/10.1016/j.cej.2023.143283.

[12] Trench AB, Moura JPC, Antonin VS, Gentil TC, Lanza MRV, Santos MC. Using a novel gas diffusion electrode based on Vulcan XC-72 carbon modified with Nb2O5 nanorods for enhancing H2O2 electrogeneration. Journal of Electroanalytical Chemistry 2023;946. https://doi.org/10.1016/j.jelechem.2023.117732.

[13] Antonin VS, Lucchetti LEB, Souza FM, Pinheiro VS, Moura JPC, Trench AB, et al. Sodium niobate microcubes decorated with ceria nanorods for hydrogen peroxide electrogeneration: An experimental and theoretical study. J Alloys Compd 2023;965:171363. https://doi.org/10.1016/j.jallcom.2023.171363.

[14] Yu Y, Zeng W, Zhang H. Hydrothermal synthesis of assembled WO3·H2O nanoflowers with enhanced gas sensing performance. Mater Lett 2016;171:162–5. https://doi.org/10.1016/j.matlet.2016.02.077.

[15] Cao S, Chen H. Nanorods assembled hierarchical urchin-like WO3 nanostructures: Hydrothermal synthesis, characterization, and their gas sensing properties. J Alloys Compd 2017;702:644–8. https://doi.org/10.1016/j.jallcom.2017.01.232.

[16] Yao S, Qu F, Wang G, Wu X. Facile hydrothermal synthesis of WO3 nanorods for photocatalysts and supercapacitors. J Alloys Compd 2017;724:695–702. https://doi.org/10.1016/j.jallcom.2017.07.123.

[17] Rojas-Carbonell S, Artyushkova K, Serov A, Santoro C, Matanovic I, Atanassov P. Effect of pH on the Activity of Platinum Group Metal-Free Catalysts in Oxygen Reduction Reaction. ACS Catal 2018;8:3041–53. https://doi.org/10.1021/acscatal.7b03991.

[18] Wan K, Yu ZP, Li XH, Liu MY, Yang G, Piao JH, et al. pH Effect on Electrochemistry of Nitrogen-Doped Carbon Catalyst for Oxygen Reduction Reaction. ACS Catal 2015;5:4325–32. https://doi.org/10.1021/acscatal.5b01089.

[19] Chen M, Ping Y, Li Y, Cheng T. Insights into the pH-dependent Behavior of N-Doped Carbons for the Oxygen Reduction Reaction by First-Principles Calculations. Journal of Physical Chemistry C 2021;125:26429–36. https://doi.org/10.1021/acs.jpcc.1c07362.

[20] Li S, Xiao C, Jiang H, Li Y, Li C. Effects of electrolytes on two-electron ORR single-atom catalysis. Sci Bull (Beijing) 2023;68:25–8. https://doi.org/10.1016/j.scib.2022.12.019.

[21] Ruggiero BN, Sanroman Gutierrez KM, George JD, Mangan NM, Notestein JM, Seitz LC. Probing the relationship between bulk and local environments to understand impacts on electrocatalytic oxygen reduction reaction. J Catal 2022;414:33–43. https://doi.org/10.1016/j.jcat.2022.08.025.

[22] Perego C, Villa P. Catalyst preparation methods. vol. 34. 1997.

[23] Moura JPC, Antonin VS, Trench AB, Santos MC. Hydrogen peroxide electrosynthesis: A comparative study employing Vulcan carbon modification by different $MnO_2$ nanostructures. Electrochim Acta 2023;463. https://doi.org/10.1016/j.electacta.2023.142852.

[24] Giannozzi P, Baroni S, Bonini N, Calandra M, Car R, Cavazzoni C, et al. QUANTUM ESPRESSO: A modular and open-source software project for quantum simulations of materials. Journal of Physics Condensed Matter 2009;21. https://doi.org/10.1088/0953-8984/21/39/395502.

[25] Prandini G, Marrazzo A, Castelli IE, Mounet N, Marzari N. Precision and efficiency in solid-state pseudopotential calculations. NPJ Comput Mater 2018;4. https://doi.org/10.1038/s41524-018-0127-2.

[26] Perdew JP, Burke K, Ernzerhof M. Generalized Gradient Approximation Made Simple. 1996.

[27] Cococcioni M, De Gironcoli S. Linear response approach to the calculation of the effective interaction parameters in the LDA+U method. Phys Rev B Condens Matter Mater Phys 2005;71. https://doi.org/10.1103/PhysRevB.71.035105.

[28] Nørskov JK, Rossmeisl J, Logadottir A, Lindqvist L, Kitchin JR, Bligaard T, et al. Origin of the overpotential for oxygen reduction at a fuel-cell cathode. Journal of Physical Chemistry B 2004;108:17886–92. https://doi.org/10.1021/jp047349j.

[29] Li G, Wang Y, Bi J, Huang X, Mao Y, Luo L, et al. Partial Oxidation Strategy to Synthesize $WS_2/WO_3$ Heterostructure with Enhanced Adsorption Performance for Organic Dyes: Synthesis, Modelling, and Mechanism. Nanomaterials 2020, Vol 10, Page 278 2020;10:278. https://doi.org/10.3390/NANO10020278.

[30] Mehta SS, Nadargi DY, Tamboli MS, Alshahrani T, Minnam Reddy VR, Kim ES, et al. $RGO/WO_3$ hierarchical architectures for improved $H_2S$ sensing and highly efficient solar-

driving photo-degradation of RhB dye. Scientific Reports 2021 11:1 2021;11:1–17. https://doi.org/10.1038/s41598-021-84416-1.

[31] Minh Vuong N, Kim D, Kim H. Porous Au-embedded WO3 Nanowire Structure for Efficient Detection of CH4 and H2S. Scientific Reports 2015 5:1 2015;5:1–13. https://doi.org/10.1038/srep11040.

[32] Zhang C, Shen W, Guo K, Xiong M, Zhang J, Lu X. A Pentagonal Defect-Rich Metal-Free Carbon Electrocatalyst for Boosting Acidic O2Reduction to H2O2Production. J Am Chem Soc 2023;145:11589–98. https://doi.org/10.1021/JACS.3C00689.

[33] Yuan J, Yin H, Ge X, Pan R, Huang C, Chen D, et al. Superior efficiency hydrogen peroxide production in acidic media through epoxy group adjacent to Co-O/C active centers on carbon black. Chemical Engineering Journal 2023;465:1385–8947. https://doi.org/10.1016/j.cej.2023.142691.

[34] Lu Z, Chen G, Siahrostami S, Chen Z, Liu K, Xie J, et al. High-efficiency oxygen reduction to hydrogen peroxide catalysed by oxidized carbon materials. Nat Catal 2018;1:156–62. https://doi.org/10.1038/s41929-017-0017-x.

[35] Chu L, Sun Z, Cang L, Wang X, Fang G, Gao J. Identifying the roles of oxygen-containing functional groups in carbon materials for electrochemical synthesis of H2O2. J Environ Chem Eng 2023;11. https://doi.org/10.1016/j.jece.2023.109826.

[36] Yuan Y, Lee TR. Contact angle and wetting properties. Springer Series in Surface Sciences 2013;51:3–34. https://doi.org/10.1007/978-3-642-34243-1_1/FIGURES/16.

[37] Kung CH, Sow PK, Zahiri B, Mérida W. Assessment and Interpretation of Surface Wettability Based on Sessile Droplet Contact Angle Measurement: Challenges and Opportunities. Adv Mater Interfaces 2019;6:1900839. https://doi.org/10.1002/ADMI.201900839.

[38] Dong H, Dong B, Sun L, Chi Z, Wang M, Yu H. Electro-UV/H2O2 system with RGO-modified air diffusion cathode for simulative antibiotic-manufacture effluent treatment. Chemical Engineering Journal 2020;390. https://doi.org/10.1016/j.cej.2020.124650.

[39] Li L, Bai J, Chen S, Zhang Y, Li J, Zhou T, et al. Enhanced O2−[rad] and HO[rad] via in situ generating H2O2 at activated graphite felt cathode for efficient photocatalytic fuel cell. Chemical Engineering Journal 2020;399. https://doi.org/10.1016/j.cej.2020.125839.

[40] Li L, Li J, Fang F, Zhang Y, Zhou T, Zhou C, et al. Efficient H2O2 production from urine treatment based on a self-biased WO3/TiO2-Si PVC photoanode and a WO3/CMK-3 cathode. Appl Catal B 2023;333. https://doi.org/10.1016/j.apcatb.2023.122776.

[41] Choi S, Lee SK, Kim NH, Kim S, Lee YN. Raman Spectroscopy Detects Amorphous Carbon in an Enigmatic Egg From the Upper Cretaceous Wido Volcanics of South Korea. Front Earth Sci (Lausanne) 2020;7. https://doi.org/10.3389/feart.2019.00349.

[42] Cuesta,'' A, Dhamelincourt P, Laureyns ' J, Martinez-Alonso ' A, Tasc~)n' JMD. RAMAN MICROPROBE STUDIES ON CARBON MATERIALS. vol. 32. 1994.

[43] Henry DG, Jarvis I, Gillmore G, Stephenson M. Raman spectroscopy as a tool to determine the thermal maturity of organic matter: Application to sedimentary,

metamorphic and structural geology. Earth Sci Rev 2019;198. https://doi.org/10.1016/j.earscirev.2019.102936.

[44] Zhang Y, Daniel G, Lanzalaco S, Isse AA, Facchin A, Wang A, et al. H 2 O 2 production at gas-diffusion cathodes made from agarose-derived carbons with different textural properties for acebutolol degradation in chloride media 2021. https://doi.org/10.1016/j.jhazmat.2021.127005.

[45] Jensen KD, Tymoczko J, Rossmeisl J, Bandarenka AS, Chorkendorff I, Escudero-Escribano M, et al. Elucidation of the Oxygen Reduction Volcano in Alkaline Media using a Copper–Platinum(111) Alloy. Angewandte Chemie 2018;130:2850–5. https://doi.org/10.1002/ange.201711858.

[46] Yang H, An N, Kang Z, Menezes PW, Chen Z. Understanding Advanced Transition Metal-based Two Electron Oxygen Reduction Electrocatalysts from the Perspective of Phase Engineering. Advanced Materials 2024. https://doi.org/10.1002/adma.202400140.

[47] Kelly SR, Kirk C, Chan K, Nørskov JK. Electric field effects in oxygen reduction kinetics: rationalizing ph dependence at the pt(111), au(111), and au(100) electrodes. AIChE Annual Meeting, Conference Proceedings 2020;2020-November:14581–91. https://doi.org/10.1021/ACS.JPCC.0C02127.

[48] Eckardt M, Sakaushi K, Lyalin A, Wassner M, Hüsing N, Taketsugu T, et al. The role of nitrogen-doping and the effect of the pH on the oxygen reduction reaction on highly active nitrided carbon sphere catalysts. Electrochim Acta 2019;299:736–48. https://doi.org/10.1016/j.electacta.2019.01.046.

[49] Mei D, He Z Da, Zheng YL, Jiang DC, Chen YX. Mechanistic and kinetic implications on the ORR on a Au(100) electrode: PH, temperature and H-D kinetic isotope effects. Physical Chemistry Chemical Physics 2014;16:13762–73. https://doi.org/10.1039/c4cp00257a.

[50] Li MF, Liao LW, Yuan DF, Mei D, Chen YX. PH effect on oxygen reduction reaction at Pt(1 1 1) electrode. Electrochim Acta 2013;110:780–9. https://doi.org/10.1016/j.electacta.2013.04.096.

[51] Koper MTM. Theory of multiple proton-electron transfer reactions and its implications for electrocatalysis. Chem Sci 2013;4:2710–23. https://doi.org/10.1039/c3sc50205h.

[52] Tse ECM, Barile CJ, Kirchschlager NA, Li Y, Gewargis JP, Zimmerman SC, et al. Proton transfer dynamics control the mechanism of O2 reduction by a non-precious metal electrocatalyst. Nat Mater 2016;15:754–9. https://doi.org/10.1038/nmat4636.

[53] Wan K, Yu ZP, Li XH, Liu MY, Yang G, Piao JH, et al. pH Effect on Electrochemistry of Nitrogen-Doped Carbon Catalyst for Oxygen Reduction Reaction. ACS Catal 2015;5:4325–32. https://doi.org/10.1021/acscatal.5b01089.

[54] Mi X, Han J, Sun Y, Li Y, Hu W, Zhan S. Enhanced catalytic degradation by using RGO-Ce/WO3 nanosheets modified CF as electro-Fenton cathode: Influence factors, reaction mechanism and pathways. J Hazard Mater 2019;367:365–74. https://doi.org/10.1016/j.jhazmat.2018.12.074.

[55] Hernández-Ferrer J, Gracia-Martín M, Benito AM, Maser WK, García-Bordejé E. Effect of temperature and presence of minor amount of metal on porous carbon materials derived from ZIF8 pyrolysis for electrocatalysis. Catal Today 2023. https://doi.org/10.1016/j.cattod.2022.12.024.

[56] Pinheiro VS, Paz EC, Aveiro LR, Parreira LS, Souza FM, Camargo PHC, et al. Mineralization of paracetamol using a gas diffusion electrode modified with ceria high aspect ratio nanostructures. Electrochim Acta 2019;295:39–49. https://doi.org/10.1016/j.electacta.2018.10.097.

[57] Valim RB, Carneiro JF, Lourenço JC, Hammer P, dos Santos MC, Rodrigues LA, et al. Synthesis of Nb2O5/C for H2O2 electrogeneration and its application for the degradation of levofloxacin. J Appl Electrochem 2023;1:1–15. https://doi.org/10.1007/S10800-023-01975-Z/FIGURES/7.

[58] Moreira FC, Boaventura RAR, Brillas E, Vilar VJP. Electrochemical advanced oxidation processes: A review on their application to synthetic and real wastewaters. Appl Catal B 2017;202:217–61. https://doi.org/10.1016/J.APCATB.2016.08.037.

[59] Wang J, Li C, Rauf M, Luo H, Sun X, Jiang Y. Gas diffusion electrodes for H2O2 production and their applications for electrochemical degradation of organic pollutants in water: A review. Science of The Total Environment 2021;759:143459. https://doi.org/10.1016/J.SCITOTENV.2020.143459.

[60] Kinoshita K (Kim), Electrochemical Society. Electrochemical oxygen technology 1992:431.

[61] Ranganathan S, Sieber V. Recent advances in the direct synthesis of hydrogen peroxide using chemical catalysis—a review. Catalysts 2018;8. https://doi.org/10.3390/catal8090379.

[62] Siahrostami S, Verdaguer-Casadevall A, Karamad M, Deiana D, Malacrida P, Wickman B, et al. Enabling direct H2O2 production through rational electrocatalyst design. Nat Mater 2013;12:1137–43. https://doi.org/10.1038/nmat3795.

[63] Kulkarni A, Siahrostami S, Patel A, Nørskov JK. Understanding Catalytic Activity Trends in the Oxygen Reduction Reaction. Chem Rev 2018;118:2302–12. https://doi.org/10.1021/acs.chemrev.7b00488.

[64] Hurtado-Aular O, Vidal AB, Sierraalta A, Añez R. Periodic DFT study of water adsorption on m-WO3(001), m-WO3(100), h-WO3(001) and h-WO3(100). Role of hydroxyl groups on the stability of polar hexagonal surfaces. Surf Sci 2020;694. https://doi.org/10.1016/j.susc.2019.121558.

[65] Qin Y, Liu M, Ye Z. A DFT study on WO3 nanowires with different orientations for NO2 sensing application. J Mol Struct 2014;1076:546–53. https://doi.org/10.1016/j.molstruc.2014.08.034.

[66] Kang M, Liang J, Wang F, Chen X, Lu Y, Zhang J. Structural design of hexagonal/monoclinic WO3 phase junction for photocatalytic degradation. Mater Res Bull 2020;121. https://doi.org/10.1016/j.materresbull.2019.110614.

[67]  Fazio G, Ferrighi L, Di Valentin C. Boron-doped graphene as active electrocatalyst for oxygen reduction reaction at a fuel-cell cathode. J Catal 2014;318:203–10. https://doi.org/10.1016/j.jcat.2014.07.024.

[68]  Heine A. Hansen; Jan Rossmeisl; Jens K. Nørskov. Surface Pourbaix diagrams and oxygenreduction activity of Pt, Ag and Ni(111) surfaces studied by DFT. Physical Chemistry Chemical Physics 2008;10:3722–30. https://doi.org/10.1039/b808799g.

Appendix A. Supplementary data

The following are the Supplementary data to this article: